\documentclass{article}
\usepackage[utf8]{inputenc}
\usepackage{amsmath} 
\usepackage{graphicx}
\usepackage{xcolor}
\usepackage{hyperref}
\usepackage{float}
\usepackage{authblk}
\usepackage{caption}
\usepackage{subcaption}

\usepackage[a4paper, total={6in, 8in}]{geometry}
\title{Methodology for physics-informed generation of \\ synthetic neutron time-of-flight measurement data}

\author[1]{Noah Walton}
\author[2]{Jesse Brown}
\author[1]{William Fritsch}
\author[3]{Dave Brown}
\author[3]{Gustavo Nobre}
\author[1]{Vladimir Sobes}

\affil[1]{University of Tennessee}
\affil[2]{Oak Ridge National Laboratory}
\affil[3]{Brookhaven National Laboratory}

\begin{document}

\maketitle

\begin{abstract}
    Accurate neutron cross section data are a vital input to the simulation of nuclear systems for a wide range of applications from energy production to national security.
    The evaluation of experimental data is a key step in producing accurate cross sections.
    There is a widely recognized lack of reproducibility in the evaluation process due to its artisanal nature and therefore there is a call for improvement within the nuclear data community.
    This can be realized by automating/standardizing viable parts of the process, namely, parameter estimation by fitting theoretical models to experimental data.
    There are numerous candidate methods to approach this type of problem, but many rely on large, labelled datasets that are not accessible to the nuclear data evaluator.
    For a reaction cross-section, there are usually just a handful of datasets, none of which can be considered labelled because evaluators never have access to the exact solution (cross section).
    This work leverages problem-specific physics, Monte Carlo sampling, and a general methodology for data synthesis to generate unlimited, labelled experimental cross-section data of high-utility.
    The synthesized data is said to be of high-utility because it is statistically similar to the observed data.
    Heuristic and, where applicable, rigorous statistical comparisons to observed data support this claim.
    The methodology is split into two generative models.
    The first generates a realization of an energy-differential cross section for a given isotope.
    The second takes the output from the first as a determined input and generates noisy experimental observables (radiation detector signals) from the determined cross section realization.
    The latter is the primary development of this article and is based/limited to transmission measurements at Rensselaer Polytechnic Institute (RPI).
    The former leverages an existing method for model parameter sampling in the resolved resonance region (RRR), thus limiting the current demonstration to the RRR of incident neutron energies.
    An open-source software is published alongside this article that executes the complete methodology to produce high-utility synthetic datasets.
    The goal of this work is to provide an approach and corresponding tool that will allow the evaluation community to begin exploring more data-driven, ML-based solutions to long-standing challenges in the field.
\end{abstract}

\section{Introduction}

A primary objective of the nuclear data community of scientists is to produce recommended mean values and covariance for neutron induced reaction cross sections.
Mean values for reaction cross sections are a fundamental input to predictive modelling and simulation of nuclear systems.
These predictive simulations underpin many applications in fields of both engineering and fundamental science (i.e., astrophysics, power-reactor operation, stockpile stewardship, non-proliferation missions, and medical isotope production for treatment or imaging).
Additionally, the uncertainty, or covariance, associated with the mean values can be systematically propagated to the simulated output quantities.
This allows for construction of confidence intervals on predicted quantities, a major consideration when making engineering or safety decisions based on predictive modelling.
Nuclear data evaluation is an integral step in obtaining accurate reaction cross sections (mean/covariance).
Recently, the community has recognized a need to improve reproducibility in the evaluation process.
The importance of this is formally addressed by the Working Party on International Nuclear Data Evaluation Co-operation (WPEC), subgroup 49 -- Reproducibility in Nuclear Data Evaluation \cite{WPECSG49}.

The need for reproducibility in the evaluation process can be addressed through automation.
The expertise of the evaluator will always be needed, but introducing automation, where-possible, can provide a systematic and reproducible starting point.
For example, a significant effort is the analysis of energy-differential experimental cross section data which would be a prime candidate for automation.
In this context, automation implies a standardized computational methodology or set of procedures that will make it easier to document artisanal decisions made by the evaluator.
Some efforts have been made on this front in the form of data/workflow \cite{SCHNABEL2021239} or of entire reaction modelling codes such as TALYS \cite{TALYS_2023}.
Access to unlimited, labelled data would be a valuable resource for developing and quantitatively benchmarking such standardized computational methodologies.
Here, labelled data refers to data sets where exact solutions are known and accessible to the computational method.
Differential cross section data -- for a single reaction with a single isotope -- is not abundant and can never be considered labelled.
This article details the development of a methodology for synthetic data generation to provide such a resource to be used in the development of new and existing computational methodologies for nuclear data evaluation.
Also, an associated open-source software to execute this methodology and produce synthetic data is published alongside this article.

Leveraging synthetic data generation methodologies in this manner has seen success in a number of ML and statistical applications (discussed in section \ref{sec:general_syndat}).
Additionally, when training a model to estimate uncertainty, it allows for a frequentist verification of the uncertainty quantification (UQ).
A relevant example of this is in \cite{refId0} where researchers used `fake' experimental cross section data to verify and compare classical and Monte Carlo (MC) Bayesian methods for uncertainty estimation.

This article develops a generative model for the experimental observables produced by a determined total cross section in a neutron time-of-flight (TOF) transmission experiment.
The TOF experiment is the primary method by which energy-differential cross sections are measured and a transmission measurement is the primary measurement used to infer total cross section.
This generative model takes a determined total cross section as an input.
As a consequence, a second generative model is needed to obtain realizations of that determined total cross section.
The framework of the former generative model (section \ref{sec:genmod2}) could be applied to TOF transmission data generally while the latter is only developed for resolved resonance range (RRR) total cross sections (section \ref{sec:genmod1}).

A demonstration of this methodology on RRR cross sections specifically is supported in a few ways.
Firstly, many applications are particularly sensitive to the accuracy and uncertainty of cross sections in this energy regime.
Also, accurately characterizing this region is foundational to the evaluation of other energy regimes.
The importance of RRR evaluations is detailed in a number of comprehensive reviews \cite{JeffReport18}\cite{AtlasofResonances}.
Secondly, the need for reproducibility in RRR evaluations is pertinent.
The default method for fitting to experimental data -- a Bayesian method which reduces to generalized linear least squares (GLLS) under certain conditions -- suffers a number of shortcomings in practice.
1) It does not consider discrete variables or parameter cardinality and 2) it is well documented that the uncertainty quantification (UQ) is underpredictive \cite{sammy}\cite{USU_NuclearData}, the presumed cause being the violation of assumptions in the GLLS regression.
Some efforts have been made to either subvert these assumptions \cite{MCMC_bayes} or meet them \cite{NDTemplates}.
However, a widely accepted, systematic solution has not be reached, resulting in significant disagreement on evaluated values and an overall lack of reproducibility in the process.
A recent proof-of-concept \cite{ND2022_Walton} showcases the potential of using synthetic data to explore novel, data-driven methods to address these challenges in RRR evaluations.
Expanding upon the referenced work can be considered near-term future work associated with this article.
Lastly, the phenomenological nature of the theoretical model in the RRR, R-matrix theory \cite{RMatrix_theory},
lends itself to synthetic data generation in that it is well-established and, if all resonances are exactly characterized, model defects are often negligible (further discussion in \ref{sec:RRR}).


\section{Methodology}


\subsection{General Data Generation}\label{sec:general_syndat}

Generating synthetic data is a common practice for many ML or statistical applications (e.g., healthcare analysis \cite{Synthetic_CTCovid_data_2022}\cite{synthetic_data_medicalML}, computer vision \cite{9511233}, software development \cite{whiting2008creating}). The general goal for many of these applications is to synthesize data that mimics the statistical properties of the observed data. The utility of the synthesized data is a measure of how well the statistical properties of the synthetic data match that of the observed data. A higher utility results in a better-trained model, a more accurate performance assessment, and higher confidence when applying trained methods to real data sets. The general process is to model the observed data (or the process by which it is generated) and then statistically sample realizations from that model to create a synthesized data set:

\begin{enumerate}
    \item Assume some joint generative distribution, parameterized by $\theta$, from which both observed data, $X$, and synthetic data, $Y$, are generated:
          \begin{equation} X,Y \sim f(\theta)\end{equation}
    \item Estimate $\theta$ to approximate a distribution for $Y$:
          \begin{equation}\hat{\theta}\ \tilde{=}\ \theta\end{equation}
    \item Sample from the approximated generative distribution:
          \begin{equation}f(Y | X, \hat{\theta})\end{equation}
    \item The utility of the synthesized data describes how well the underlying distribution is approximated:
          \begin{equation}f(Y | X, \hat{\theta}) \sim f(\theta)\end{equation}
\end{enumerate}

Step 2 is the most application-specific; there are many ways to attempt to characterize statistical distributions of observed data. The model and its parameters, $f(\theta)$, can be inferred using ML methods with no underlying assumptions or it can be completely physics-informed, not reliant on any observed data. Integration of a physics-based model can greatly improve utility while reducing the need for copious amounts of observed, labelled data. This method has been employed for manufacturing processes \cite{en15145085}\cite{BatteryManufacturing_syntheticData}, biological experiment design \cite{Tcell_SyntheticData}, and nuclear non-proliferation applications \cite{Radio_isotope_identification}\cite{9121756}. When using a well-established, physics-based model, step 3 often takes the form of uncertainty sampling and/or noise sampling. If a process is well described by a physical model, then the statistical variation in the data must be a function of input parameter uncertainty or noise in the observables.


\subsection{Application to Resolved Resonance Range Data}
\label{sec:RRR}

In the case of RRR cross sections, the theoretical model is given by R-Matrix theory \cite{RMatrix_theory} and the model parameters can be called resonance parameters.
The phenomenological nature of this model is uniquely advantageous in that, if the resonance parameters are perfectly described, model defects in the calculation of the cross section are minimal.
This is especially the case for simpler isotopes (no fission channels) in the medium-to-large mass range.
This is part of the motivation for demonstrating this methodology on Ta-181.
Combining the phenomenological nature of the reaction model, an understanding of the experimental process/uncertainties, and a known noise model for observables, high-utility synthetic data can be generated with no black-box ML method.

This methodology can be separated into three major parts.
The first is a generative model for the resonance parameters.
The second is a generative model for the process by which a set of determined resonance parameters produce observable data. Lastly, a noise model is applied to the generated observable data (the determined output of the second generative model).
Note that the current methodology is developed for the total reaction cross section, which corresponds to transmission (see section \ref{sec:reduction}).

\begin{figure}[H]
    \centering
    \includegraphics[width=6in]{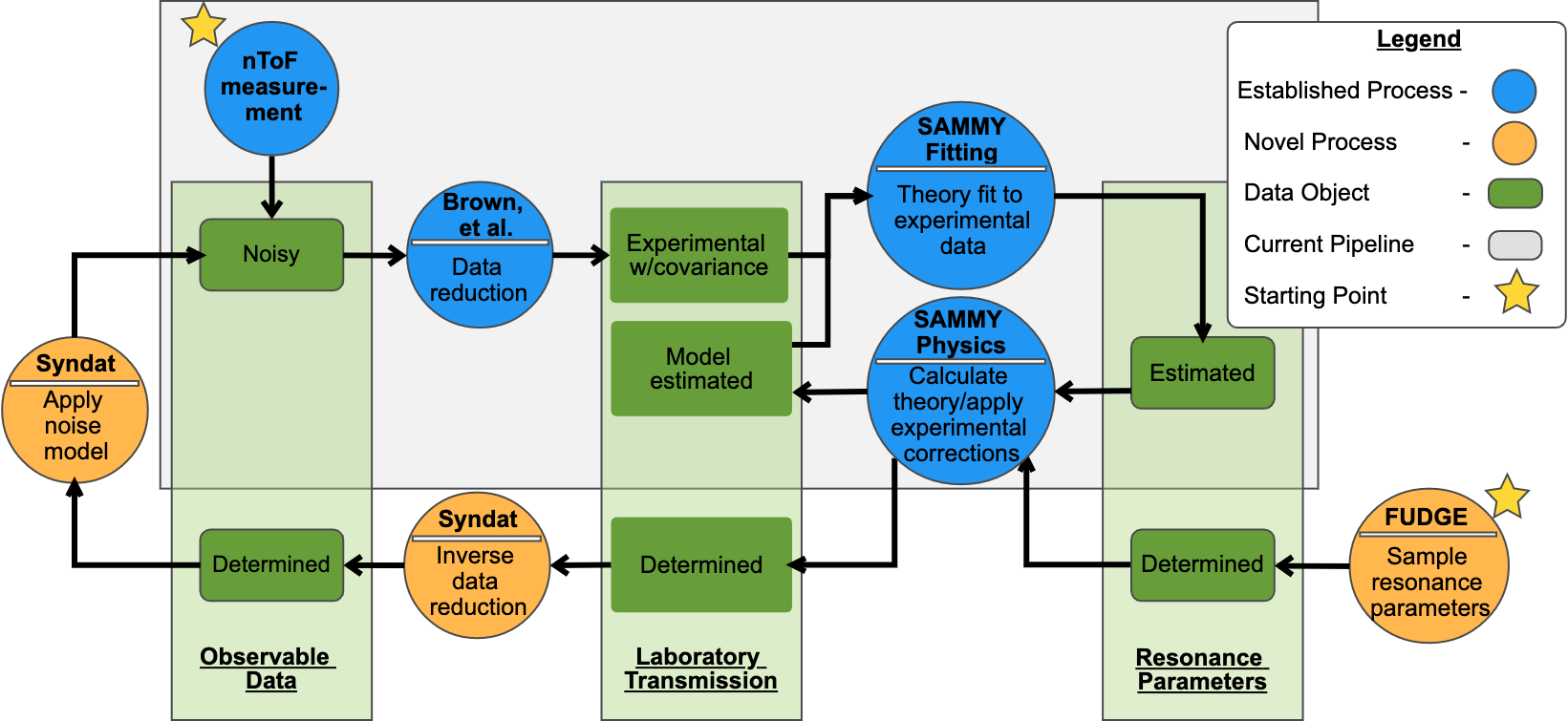}
    \caption{Data flow diagram for the data synthesis methodology. Circles represent processes, rectangles represent data. The gray border encompasses the current processes for experimental measurement and evaluation, within which established methodologies and software can be leveraged. The processes outside of this border represent the novel methods developed in this article. Data objects that are part of the data generation flow are labelled determined because the true/exact value is known.}
    \label{fig:DFD}
\end{figure}

The data flow diagram in figure \ref{fig:DFD} shows a high level overview of the data and process flow for this generation methodology. When possible, established methods and/or software is used. This is dilineated in the figure; processes that leverage established methods (and software) are shown in blue while processes that involve a novel method (and software) are shown in orange. The data-flow that falls within the gray border represents the current process for experimental measurement and evaluation. In an evaluation, the laboratory transmission data object is where theory is fit to experimental data in an attempt to determine the best resonance parameters. Notice that this object is neither directly measured nor directly calculated from theory. Rather, the data reduction process is applied to observable data and experimental corrections are applied to theory in order to reach this point of comparison.

The three major parts of the data generation methodology can be seen in the data flow diagram in orange. The first generative model for resonance parameter sampling is based on the statistical theory of resonances. This generative model has been developed in the processing code FUDGE \cite{FUDGE} to train ML models to characterize quantum spin group assignment \cite{ML_spingroups} and is briefly discussed in section \ref{sec:genmod1}.
The second generative model uses established methods for calculating an experimentally corrected theoretical model (to get the laboratory transmission data object) and a novel method for inverse data reduction to generate observables. This is primary focus of the work presented in this article. The methodology is discussed in \ref{sec:genmod2} and is executed using the SAMMY code developed at Oak Ridge National Laboratory (ORNL) \cite{sammy} and the open-source code published alongside this article, Syndat (linked in the conclusion of this article).
This generative model is highly specific to the experimental parameters and procedures. Here, it is developed using the forward reduction methodology for transmission measurements in \cite{BrownThesis} which correspond to the total reaction cross section for Ta-181. This reduction methodology is summarized in section \ref{sec:reduction} as it is useful for understanding the inverse methodology. While a general set of experimental parameters and/or procedure may suffice, the synthesized data will be more representative of the observed data set if both are generated from similar experimental conditions. The extension of this methodology to include other reaction measurements and default experimental models can be considered future work.


\subsection{Reduction methodology}\label{sec:reduction}

As aforementioned, the second generative model for this data generation methodology is dependent heavily on the experiment and data reduction processes.
The experiment and reduction methodology on which this work is based is described in detail in \cite{BrownThesis}. The referenced work is associated with a set of transmission measurements for $^{181}$Ta at the Rensselaer Gaerttner Linear Accelerator, one of the few facilities in the world where RRR differential measurements can be performed.
Transmission measurements are used to infer total reaction cross sections, the relationship between the two is described in section \ref{sec:theo_trans}.
The following equations describe the data reduction process used to calculate transmission from a set of observables (detector counts).

\begin{equation}\label{eq:trans_exp}
    T = \frac{\text{sample\ in}}{\text{sample\ out}} =
    \frac{ \alpha_1\dot{c}_\text{s} - \alpha_2 k_\text{s} \dot{B} - \dot{B0}_\text{s} }
    {\alpha_3\dot{c}_\text{o} - \alpha_4 k_\text{o} \dot{B} - \dot{B0}_\text{o}}
\end{equation}

\begin{equation}\label{eq:dtctr}
    \dot{c} = \frac{c}{b_\text{w}\, \text{trig}} \times dtcf
\end{equation}

\begin{equation}\label{eq:bkg}
    \dot{B}(t_i) = a e^{-b t_i}
\end{equation}

\begin{center}
    \begin{tabular}{ |c|l| }
        \hline
        Variable           & Description                                                   \\
        \hline
        $T$                & Transmission                                                  \\
        $\dot{c}_{s/o}$    & Dead time corrected count rates for sample in/out             \\
        $\alpha_{1,2,3,4}$ & Monitor normalization factors for each cycle (1-4)            \\
        $k_{s/o}$          & Normalization for time-dependent background for sample in/out \\
        $\dot{B}$          & Time dependent background function                            \\
        $\dot{B0}_{s/o}$   & Constant room background function for sample in/out           \\
        $c$                & Detector counts for either sample in or out                   \\
        $dtcf$             & Detector dead-time correction factor (see \cite{BrownThesis}) \\
        $b_w$              & Time bin width of the detector system                         \\
        $trig$             & Number of times that bin was open for detection               \\
        \hline
    \end{tabular}
\end{center}

It can be seen that the transmission is effectively calculated as a ratio between the number of neutrons that traverse a sample to the number of neutrons that are incident upon it. Generally, a measurement consists of a number of cycles. Each cycle produces a vector of counts dependent on time-of-flight (TOF), the notation for time dependence above has been dropped for conciseness but note that the equations above should be vectorized with elements corresponding to discrete TOF bins. The TOF bin widths for the raw detection system are on the order of 6~ns. In the reference experiment, this data is dead time corrected and monitor normalized for each cycle, then, summed over all cycles, though other methods exist for this step.
Next, in an artisanal manner, the experimentalist performing the data reduction will select some re-binning structure, increasing the number of counts per time bin and bettering counting the statistics at the cost of a decreased resolution in TOF. From here, the equations above are used to calculate a spectra of transmission with respect to TOF bins defined by the experimentalist. Additionally, uncertainties are propagated through the above process with systematic error arising from the background functions and non-systematic or statistical errors arising from detector counting statistics \cite{BrownThesis}\cite{tsoulfanidis2015measurement}.


\subsection{Generative Model 1: Resonance Parameters}
\label{sec:genmod1}

This section describes the methodology for sampling a set of resonance parameters corresponding to a theoretical cross section.
This will be the ``solution" to the synthesized experimental data.
The concept of sampling resonance parameters is supported by the statistical theory of neutron resonances.
Resonance structures in neutron cross sections correspond to excited states of the compound nucleus formed when an incident neutron is absorbed and equilibrates in a target nucleus.
Each of these levels belongs to a discrete quantum spin group, characterized by the orbital angular momentum and parity of the particle-pair.
For any given spin group, resonance parameters are observed to fall on well-characterized statistical distributions.
These statistical distributions provide the basis for a generative probability model.

Resonance structures are parameterized by a location in energy and a total width made up of constituent reaction-channel widths.
For any given spin group, single-channel widths (i.e., elastic scatter) tend to fall on the Porter-Thomas distribution, a \begin{math}\chi ^2\end{math} distribution with one degree of freedom.
Reaction widths with more than one channel are described by a \begin{math}\chi ^2\end{math} distribution with degrees of freedom equal to the number of channels.
Similarly, the energy spacing between neighboring resonances of the same spin group tend to fall on the Wigner distribution.
This distribution can be derived from random matrix theory, particularly the Gaussian Orthogonal Ensemble (GOE), and our current understanding of quantum chaotic systems.
The probability distributions for resonance widths and spacings were first introduced in \cite{PT_original} and \cite{wigner_distr_original} respectively, but are widely used in modern evaluation techniques both in the resolved resonance region and beyond \cite{JeffReport18}\cite{TENDL_2019}\cite{EAF_2000}.
The applicability of these statistical distributions is further supported by their use in representing the unresolved resonance range (URR) in continuous energy Monte Carlo transport calculations.
The onset of the URR occurs where individual resonance structures cannot be fully resolved experimentally but still cause large variance in the mean cross section.
This energy region is evaluated as an average cross section.
When accessing this cross section for Monte Carlo neutron transport simulations, however, probability tables can be used to better represent this energy regime.
Probability tables are constructed from realizations of resonance parameter sequences sampled from the Porter Thomas and Wigner distributions.
This is done in a number of processing codes \cite{NJOYmethods_LANL}\cite{AMPX_Ptables}\cite{FUDGE}\cite{CALENDF}. This work leverages the high-fidelity implementation developed by Brookhaven National Laboratory (BNL) \cite{ResTools_BNL} distributed with the open source processing code FUDGE \cite{FUDGE} developed by Lawrence Livermore National Laboratory (LLNL).

\subsection{Generative Model 2: Physical Process}
\label{sec:genmod2}

This section describes the methodology for modelling the process by which a set of theoretical resonance parameters produce an observable. The actual observable is raw detector count data, for which there is a known noise model (Poisson counting statistics). For this section, it is assumed that the methodology in section \ref{sec:genmod1} has been used to generate a realization of resonance parameters that are taken as determined inputs to this model. In other words, this section assumes access to an exact theoretical cross section from which realizations of experimental data are generated.

The process can be split into two steps. First, the theoretical model and experimental corrections are used generate a laboratory transmission from the known resonance parameters. Second, a novel method for inverse-reduction is used to generate the expected observable (raw count data) from the laboratory transmission. Both steps require some experiment-specific parameters which are never known exactly. In all cases, a Monte Carlo sampling approach is taken to capture all possibilities within the sample space of our generative model.

\subsubsection{Experimentally observable cross section}\label{sec:theo_trans}

The goal of this section is to generate a theoretical cross section that can be observed in a laboratory setting from a set of resonance parameters and some experimental conditions. The methodology behind these experimental corrections is well established and part of the current process of evaluation (as shown in figure \ref{fig:DFD}). The reason for this direction of data/process flow in the evaluation process is due to the nature of these experimental correction. Those included in this section are characterized by the fact that the mathematical operation cannot be inverted. These operations \textit{must} be applied to the theoretical model before it can be compared to experimental data -- as opposed to applying the inverse operations to the experimental data and comparing it directly to the theoretical model.

These calculations are widely used and accepted in the evaluation process for total cross section/transmission in the RRR, often using an R-Matrix code. This work leverages the popular, open-source R-Matrix evaluation code SAMMY \cite{sammy} developed at Oak Ridge National Laboratory. The only novelty of this section is how the calculation is being used. A brief overview of this process for the transmission measurement will be given here with details left to the references \cite{sammy}\cite{BrownThesis}.

Given a set of resonance parameters, a total cross section can be calculated via R-Matrix theory \cite{RMatrix_theory}. In the case of transmission experiments, the necessary experimental corrections include only Doppler and resolution broadening which correct for the thermal motion of target particles and the uncertainty in the temporal origin of the incident particles. The nature of a transmission measurement allows for other interactions within the sample (multiple scattering phenomena) to be neglected. It is also assumed that the sample is elementally and isotopically pure, though this would be a simple addition to the methodology. The below steps show an overview of this process, in each case $f$ is meant to represent a general function with details found in the references.

\begin{enumerate}
    \item Calculate the theoretical cross section:
          \begin{equation} \sigma_t = f(resonance\ parameters) \end{equation}
    \item Doppler broaden the theoretical cross section:
          \begin{equation} \label{eq:sig_trans}\sigma_{t,D} = f(\sigma_t,temp)\end{equation}
    \item Convert the Doppler broadened cross section to transmission:
          \begin{equation}T_D = e^{-n\sigma_{t,D}}\end{equation}
    \item Resolution broaden the Doppler broadened transmission:
          \begin{equation}T_{\text{res}} = f(T_D)\end{equation}
\end{enumerate}

In the above steps for experimental correction, each correction is a function of the previously corrected object in that particular order. Note that the resolution broadening function is performed on the transmission object rather than the cross section. This results in an experimentally corrected theory that is more likely to agree with the experimental transmission measurement, the difference in methods deriving from self-shielding effects \cite{sammy}. In practice, evaluations are done using transmission, so the resolution broadened transmission ($T_{res}$) is the final product. The transmission is related to the total cross section via equation \ref{eq:sig_trans} where $n$ is the thickness of the target sample in $\text{atoms/barn}$.


\subsubsection{Novel method for inverse-reduction}\label{sec:inverse_reduction}

Leveraging the methodology in section \ref{sec:theo_trans}, an expected transmission (experimentally corrected) from sampled resonance parameters can be calculated.
Through the inversion of the reduction process described in section \ref{sec:reduction}, expected values for the observable that is driven by the expected transmission can be calculated (raw count data for sample in $c_s$).
This is shown in equations \ref{eq:cps_theo} and \ref{eq:cts_theo}, where the `theo' subscript indicates an expected value driven by the sampled resonance parameters.
To calculate $\dot{c}_{s,\mathrm{theo}}$, all other input parameters must be determined.
In actuality, the other input parameters are measured with some degree of uncertainty by the experimentalist.
These parameters are sampled for each synthetically generated dataset using the distribution estimated by the experimentalist.
This is often a normal distribution with estimated mean/variance, but is ultimately decided by the reported experimental distribution.
For example, in the case of the parameters that describe the background function ($\dot{B}$), a multivariate normal distribution with appropriate correlations is used.

\begin{equation}\label{eq:cps_theo}
    \dot{c}_{s,\mathrm{theo}} = \frac{T_{\mathrm{theo}}
        (\alpha_3\dot{c}_\text{o} - \alpha_4 k_\text{o} \dot{B} - \dot{B0}_\text{o})
        + \alpha_2 k_\text{s} \dot{B} + \dot{B0}_\text{s} }
    {\alpha_1}
\end{equation}

\begin{equation}\label{eq:cts_theo}
    c_{s,\mathrm{theo}} = \dot{c}_{s,\mathrm{theo}}(b_\text{w}\, \text{trig}) 
\end{equation}

The above process will generate samples of the \textit{expected} observable, sample-in detector counts $c_{s,\mathrm{theo}}$, as determined by the sampled resonance parameters and experimental parameters.
Here we can confidently apply the noise model to $c_{s,\mathrm{theo}}$ because radiation counting is an inherently stochastic process governed by Poisson counting statistics \cite{tsoulfanidis2015measurement}.
This distribution can be leveraged to sample statistically appropriate realizations of detector counts around the expectation value given by equation \ref{eq:cts_theo}.
At this point, we have generated a noisy, statistical realization of the observable in a neutron time-of-flight experiment.
From here, the reduction process described in section \ref{sec:reduction} can be used to reduce the synthetic data to back to transmission, the result being a noisy realization of experimental transmission data with an exact, known solution.

Two technicalities are briefly noted here.
First, the expectation values for $c_{s,\mathrm{theo}}$ are calculated using the integral of the cross section (or transmission) over the represented time bin.
This ensures that fine-structures in the cross section will influence the data even if the structures are smaller than the bin-width of the reduced data.
Secondly, because the reference experiment \cite{BrownThesis} applies monitor normalizations and dead-time corrections at each cycle, the expected count is split into multiple cycles (35 in this case) before applying the noise model.


\section{Results and Verification}

The utility of the first generative model, for resonance parameter generation, is dependent on the fidelity of average parameters and the theoretical distributions themselves.
Average parameters are inferred from evaluations, making them susceptible to bias. These can be randomly perturbed to avoid biasing the synthetic dataset to a particular set of incorrect average parameters.
The fidelity of the theory behind the resonance parameter distributions is validated by the extent to which evaluations and statistical analysis of resonances to-date follow the proposed distributions.
The caveat is that these theoretical parameter distributions are often considered in the evaluation process, making the resulting evaluation dependent on the theory.
It is beyond the scope of this article to investigate the fidelity of the statistical theory of resonances. The authors do, however, recognize that the synthetic data capability developed in this work could be leveraged for such an investigation.

The second generative model for generating noisy, experimental observables from a determined set of resonance parameters is the primary contribution of this article.
Assessing the utility of this generative model is a non-trivial task.
The primary challenge is that observed data with a labelled solution does not exist.
The reference experimental data set for $^{181}$Ta transmission can be used for heuristic comparisons, however, without access to the true underlying resonance parameters for $^{181}$Ta, a rigorous comparison is not possible.

The reference experimental data set for $^{181}$Ta transmission was used for heuristic comparison.
Using the Bayesian R-Matrix code Sammy \cite{sammy} and the $^{181}$Ta resonance evaluation in JEFF-3.3 \cite{JEFF3p3} as a prior, a GLLS fit to the reference data was found.
The resulting parameters were fixed (not sampled) and used to generate synthetic datasets per the methodology of the second generative model (Section \ref{sec:genmod2}).
The result being synthetically generated statistical realizations of experimental data from an underlying set of determined resonance parameters given by the GLLS fit.
In order to compare synthetic and observed datasets, it must be assumed that the determined resonance parameter set underlies the reference data as well.
This comparison is captured in the following figures where the assumed, underlying transmission is shown in green and the synthetically generated data can be visually compared to the reference data.
\begin{figure}[H]
    \centering
    \includegraphics[width=6in]{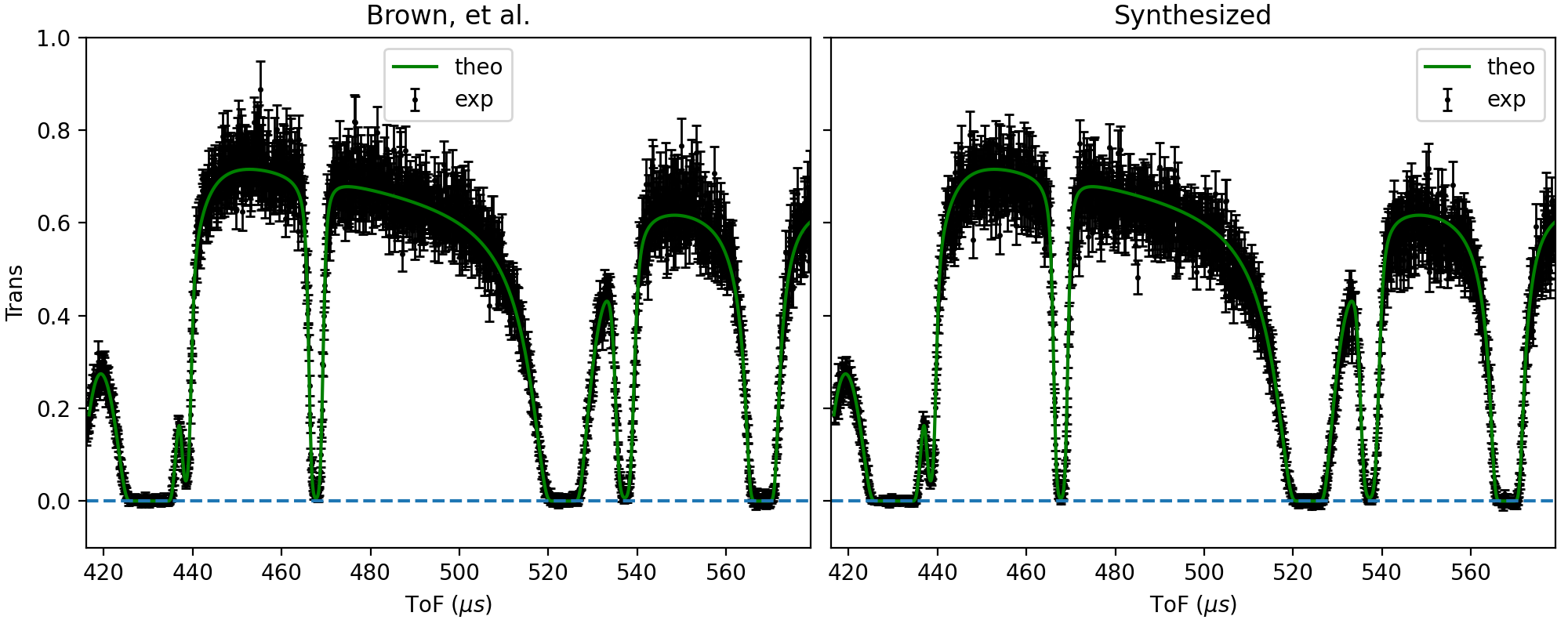}
    \caption{Observed experimental data compared to synthetically generated data assuming the GLLS solution to the observed data is the true underlying set of resonance parameters.}
    \label{fig:Syndat_vs_Brown_trans}
\end{figure}
While the above figures are meant to illustrate the likeness of observed and synthetically generated data, one intended use-case is also highlighted.
For each generated dataset, access to the underlying cross-section (in green) allows for an assessment of performance when developing an algorithm to estimate and/or provide some range of confidence on the solution.

\begin{figure}[h]
    \centering
    \includegraphics[width=15cm]{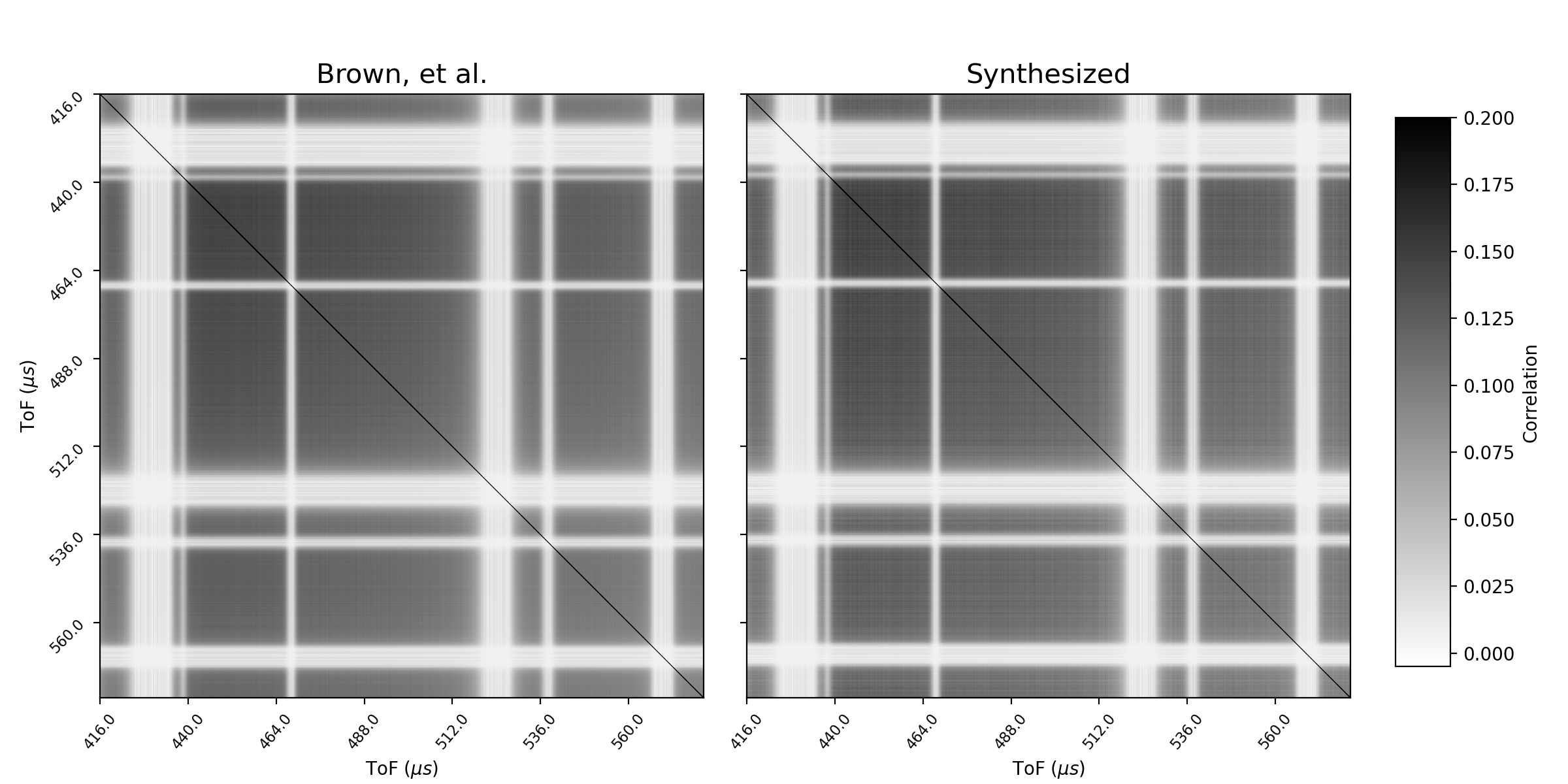}
    \caption{Time of flight correlation matrix for observed experimental data compared to synthetically generated data.}
    \label{fig:Syndat_vs_Brown_corr}
\end{figure}

Another point of comparison is the correlation structure of the uncertainty in the datapoints with respect to TOF.
Figure \ref{fig:Syndat_vs_Brown_corr} shows the correlation matrix for observed and synthetically generated data with no discernible differences in structure or magnitude.

The above figures show data generated from fixed resonance parameters such that repeated samples would produce additional statistical realizations of experimental data from the same fixed cross section -- emulating multiple experimental measurements of the exact same cross section.

The claim for a high-utility generative model is that, assuming the resonance parameters are exactly correct, the observed data is one possible realization from the ensemble distribution of synthetically generated datasets.
Any one realization can be characterized by a distribution of normalized residuals, calculated by,
\begin{equation}
    \tau = \frac{T_{\mathrm{exp}} - \mathrm{E}[T]}{\sigma_{\mathrm{exp}}}
\end{equation}
where the $T_{\mathrm{exp}}$ is the experimental data point (synthetic or observed) and $\sigma_{\mathrm{exp}}$ is its estimated standard deviation.
$\mathrm{E}[T]$ is the expectation value for the transmission which is directly related to the theoretical cross-section that produced the data.
In order to rigorously test the utility claim, access to the exact cross section that produced the observed reference data is needed.
The Porter Thomas distribution suggests that the most abundant resonances are those with a very small width.
Smaller resonances are often lost to experimental resolution and therefore missing from the evaluation and the determined cross section used to generate synthetic datasets.
Without access to the true underlying resonance parameters of $^{181}$Ta, it is likely that small resonance structures are contributing to the observed distribution.

\begin{figure}[h]
    \centering
    \begin{subfigure}[b]{0.6\textwidth}
        \centering
        \includegraphics[width=8cm]{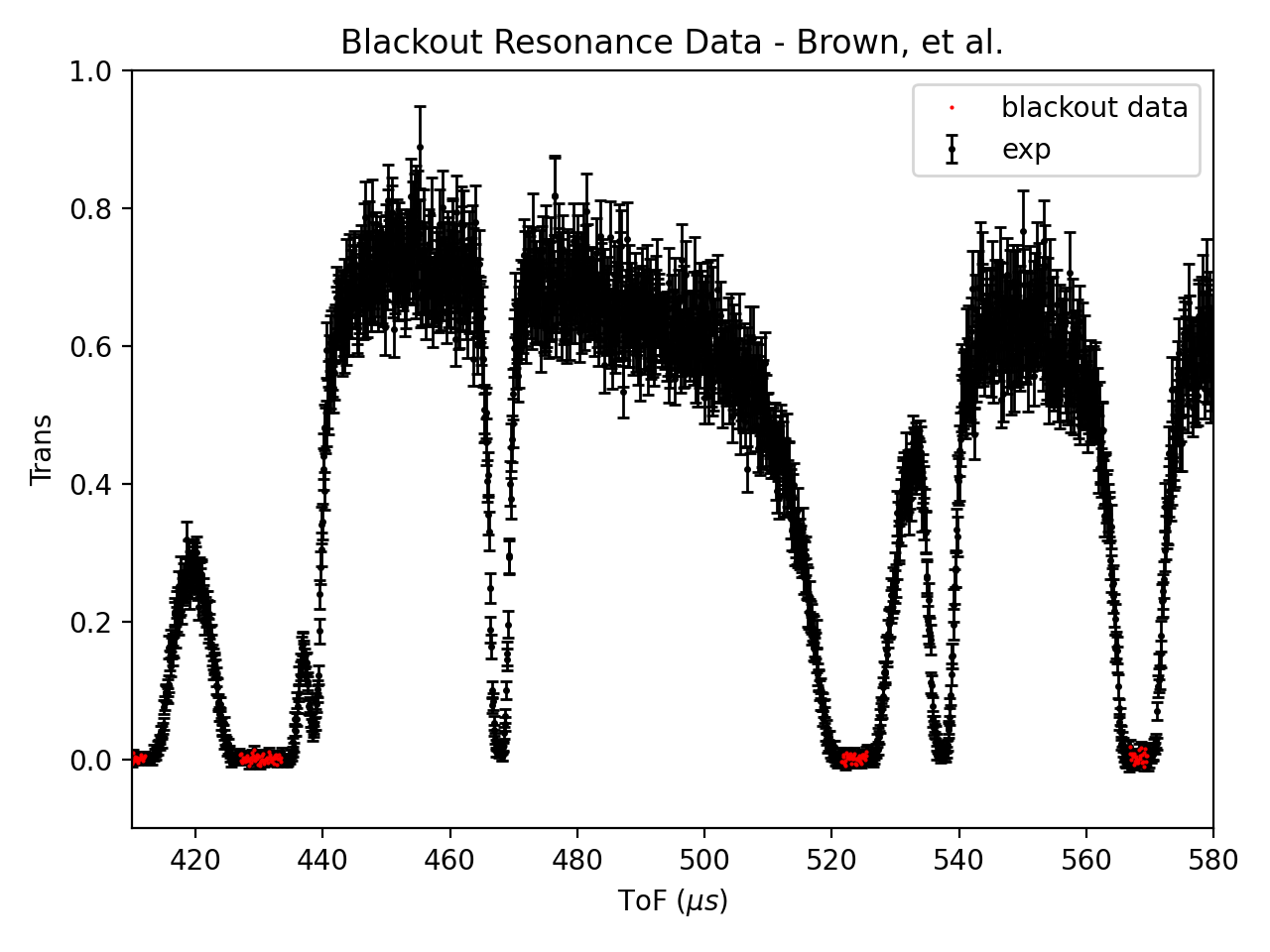}
        \caption{Blackout resonances and highlighted data where the expectation value can be well-estimated in the 410-580 $\mu s$ ToF window.}
        \label{fig:blackout_data}
    \end{subfigure}
    \begin{subfigure}[b]{0.39\textwidth}
        \centering
        \renewcommand{\arraystretch}{1.1}
        \begin{tabular}{ |c|c| }
            \hline
            Energy (eV) & ToF Range ($\mu s$) \\
            \hline
            4.280       & 1175    -   1290    \\
            10.36       & 785     -   803     \\
            13.95       & 684.5   -   688.75  \\
            20.29       & 566.75  -   569.5   \\
            23.92       & 521.5   -   525.5   \\
            35.90       & 427.0   -   433.5   \\
            39.12       & 406.75  -   412.00  \\
            99.20       & 258     -   259     \\
            174.9       & 195.1   -   195.7   \\[1ex]
            \hline
        \end{tabular}
        \vspace{0.1 cm}
        \caption{\label{tab:blackout_locations} JEFF-3.3 evaluated resonance energies nearest to the selected blackout locations and exact ToF ranges over which the expectation and residual were calculated.}
    \end{subfigure}
    \vspace{0.1 cm}
    \caption{Visualization and exact location of selected blackout resonances used in the rigorous statistical comparison of observed and synthetically generated residuals.}
    \label{fig:blackout_locations}
\end{figure}

To address this, the study can be filtered to only the `black resonances', meaning resonances that allow no neutrons to be transmitted.
At these sections of the ToF measurement, the expectation value for the transmission is known even if the theoretical cross section is not specified exactly.
Of course, the theoretical cross section has to be similar enough to have produced a black resonance in the first place.
The detector response in these ToF windows comes only from the background spectrum, therefore, the observed transmission data is a function of only the second generative model and is not affected by the potential presence of small, missed resonances in the evaluation.
This is visualized in figure \ref{fig:blackout_data}, where the blackout resonances in the 410-580 $\mu s$ ToF window are identified.
In order to improve the power of the statistical comparisons through sample size, all obviously blacked-out resonances from the RRR were considered.
The resonance energy locations that were selected are given in table \ref{tab:blackout_locations}.
The total number of blackout residuals per RRR realization (synthetic or observed) was 572.

By filtering this data to black resonances only, the data distributions are isolated to only the second generative model.
The variance of this model is driven by counting statistics and uncertainty in the data reduction process (experimental setup/parameters).
The methodology for sampling experimental parameters takes the measured value as an estimate of the mean and randomly perturbs it based on the reported uncertainty.
As a result, these estimated values drive the expectation value/variance of the ensemble distribution such that it is not directly comparable to the observed distribution.
However, the claim for high-utility synthetic data is that the ensemble distribution should fully capture the observed distribution \textbf{and} have a nonzero probability of realizing statistically identical data.
The results of this study are shown in below in figure \ref{fig:statistical_tests}.

\begin{figure}[h]
    \centering
    \begin{subfigure}[b]{0.6\textwidth}
        \centering
        \includegraphics[width=8cm]{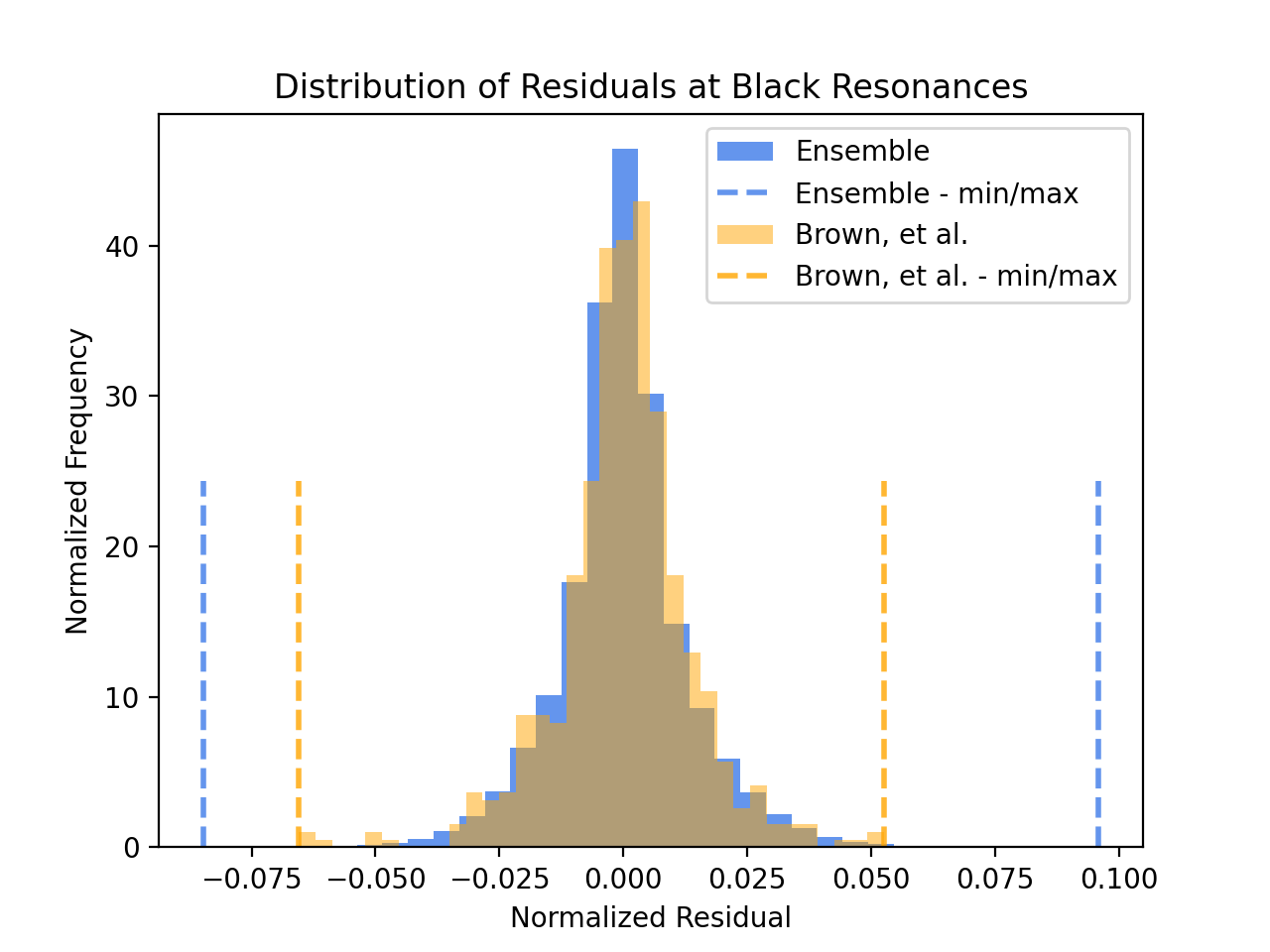}
        \caption{Histogram of observed realization and synthetic ensemble.}
        \label{fig:hist_trans_residuals_window}
    \end{subfigure}
    \begin{subfigure}[b]{0.39\textwidth}
        \centering
        \renewcommand{\arraystretch}{1.5}
        \begin{tabular}{ |c|c|c| }
            \hline
            p-value & KS test & AD test \\
            \hline
            0.05    & 54.5\%  & 58.6\%  \\
            0.10    & 47.0\%  & 49.6\%  \\
            0.15    & 41.0\%  & 41.9\%  \\
            0.20    & 36.6\%  & 35.8\%  \\
            0.25    & 32.2\%  & 32.8\%  \\
            \hline
        \end{tabular}
        \caption{\label{tab:statistical_tests} Percent realizations that failed to reject the null hypothesis for Kolmogorov-Smirnof and Anderson-Darling 2-sample statistical similarity tests at different significance levels.}
        \label{tab:statistical_passes}
        \space
    \end{subfigure}
    \vspace{0.1 cm}
    \caption{Results of the rigorous statistical comparison between observed and synthetically generated normalized residuals in the black resonance regions.}
    \label{fig:statistical_tests}
\end{figure}

The histogram (\ref{fig:hist_trans_residuals_window}) shows the ensemble distribution and the observed realization of normalized residuals.
The lines indicate the minimum and maximum of each, demonstrating that the ensemble distribution fully captures the observed realization.
For each realization that contributed to the synthetically generated ensemble distribution, statistical tests were performed to compare with the observed realization.
The two-sample Kolmogorov–Smirnov and Anderson–Darling tests \cite{KS_test}\cite{AD_test} were used to quantify the statistical similarity of the two distributions.
The first tests the null hypothesis that the two sample sets have identical underlying continuous distribution functions.
The second tests the null hypothesis that the two sample sets are drawn from the same population.
The results of these statistical similarity tests can be represented using a significance level, or p-value.
For different significance levels, the percentage of realizations for which the null hypothesis failed to be rejected is given in the table (\ref{tab:statistical_passes}).
These results show that a significant percentage of the synthetic data realizations are statistically indistinguishable from the observed data.


\section{Conclusion}

This article details the development of an approach and corresponding tool for the generation of high-utility, experimental transmission data in the resolved resonance range (RRR).
The software developed to execute this methodology was made open-source and is hosted on an online repository at \url{https://github.com/Naww137/Syndat}.
This is a multi-step process requiring: 1) a generative model for resonance parameter/cross section realizations, 2) a generative model for the process by which a cross section produces experimental observables/transmission, and 3) a noise model applied to the process model in 2.
This methodology is used to synthesize sets of experimental observables with corresponding, exact resonance parameters (labelled data).
In order to generate high-utility data, both generative models must produce datasets that are statistically indistinguishable from real or observed data.
For resonance parameter generation, the utility of the data is driven by the statistical theory of resonances and a rigorous analysis is outside the scope of this article.
For the generation of experimental observables, the methodology for high-utility synthetic data was supported by both heuristic and rigorous statistical comparisons.
This second generative model was the primary effort here and its utility is more important for future associated work.
For example, confidence in the utility of the second generative model can allow for further exploration of the statistical theory of resonances and the ability to validate it experimentally

One of the primary challenges addressed in the development of this methodology that the generative model of the process requires a detailed knowledge of the experiment being modelled.
While having access to these details is ideal for model training, hypothesizing uncertainties and/or correlations and intentionally leaving them out of the evaluation/UQ can give insight to the impact of unquantified uncertainties and the ability of the UQ to handle them.
Unquantified uncertainties and/or correlations are a major challenge in the RRR evaluation process that can limit the design, economics, and safety of many nuclear applications.
The synthetic data approach and tool developed here are resources that will allow for exploration and pseudo-validation of new and advanced methods to address these challenges.
This use-case alludes to associated future-work that will expand on the referenced proof-of-concept in \cite{ND2022_Walton}.
This work will leverage large sets of labelled data -- generated using the methodology in this article -- to develop and train an algorithm to automatically identify and characterize resonance structures in experimental data.
The accessible solutions will allow for quantitative benchmarking of the inferential capabilities and a frequentist verification of the UQ/confidence estimated by the algorithm.
The development of a systematic, automated tool for this step in RRR evaluations has the potential to greatly improve efficiency and reproducibility.
In order to integrate multiple aspects of the evaluation process into this automation effort, other experimental reaction data must be included.
Future work might consider developing parallel methodologies for generating synthetic experimental data for capture or fission time-of-flight measurements.
It should be kept in mind that the claim of statistical indistinguishability is limited by the validity of the theoretical model and should be recognized if used beyond the simpler case of Ta-181.
This article, the associated open-source software, and future-work are focused on evaluation in the RRR.
However, the framework for generating noisy experimental observables in neutron TOF experiments could be expanded to other energy regimes.
While it is not the immediate goal of the future-work associated with this article, adaptation of this methodology to other energy regimes could pose similar benefits.
Access to high-utility, labelled synthetic data will allow for a unique ability to test and benchmark the inferential limits of new and existing computational methodologies for nuclear data evaluation.


\section*{Acknowledgements}

This material is based upon work supported by the Department of Energy National Nuclear Security Administration through the Nuclear Science and Security Consortium under Award Number(s) DE-NA0003996.

This work was supported by the Nuclear Criticality Safety Program, funded and managed by the National Nuclear Security Administration for the U.S. Department of Energy. Additionally, work at Brookhaven National Laboratory was sponsored by the Office of Nuclear Physics, Office of Science of the U.S. Department of Energy under Contract No. DE-SC0012704 with Brookhaven Science Associates, LLC. This project was supported in part by the Brookhaven National Laboratory (BNL), National Nuclear Data Center under the BNL Supplemental Undergraduate Research Program (SURP) and by the U.S. Department of Energy, Office of Science, Office of Workforce Development for Teachers and Scientists (WDTS) under the Science Undergraduate Laboratory Internships Program (SULI).

This work was supported by the Nuclear Criticality Safety Program, funded and managed by
the National Nuclear Security Administration for the Department of Energy
\vskip 10pt
\noindent
This report was prepared as an account of work sponsored by an agency of the United States Government. Neither the United States Government nor any agency thereof, nor any of their employees, makes any warranty, express or implied, or assumes any legal liability or responsibility for the accuracy, completeness, or usefulness of any information, apparatus, product, or process disclosed, or represents that its use would not infringe privately owned rights. Reference herein to any specific commercial product, process, or service by trade name, trademark, manufacturer, or otherwise does not necessarily constitute or imply its endorsement, recommendation, or favoring by the United States Government or any agency thereof. The views and opinions of authors expressed herein do not necessarily state or reflect those of the United States Government or any agency thereof.

\newpage

\bibliographystyle{IEEEtran}
\bibliography{refs}

\end{document}